\documentclass[12pt]{article}
\pdfoutput=1

\PassOptionsToPackage{sort, square}{natbib}
\usepackage[preprint]{n_edited}
\usepackage{hyperref}
\hypersetup{breaklinks=true}
\usepackage{xurl}
\usepackage[hyphenbreaks]{breakurl}

\usepackage[utf8]{inputenc} 
\usepackage[T1]{fontenc}    
\PassOptionsToPackage{hyphens}{url}
\usepackage{hyperref}       
\usepackage{booktabs}       
\usepackage{amsfonts}       
\usepackage{nicefrac}       
\usepackage{microtype}      
\usepackage[dvipsnames,table,xcdraw]{xcolor}         
\usepackage{multirow}
\usepackage{nicematrix}
\usepackage{caption}
\usepackage{hhline}
\usepackage{cellspace}
\setcitestyle{numbers}

\title{International Institutions for Advanced AI}

\author{
Lewis Ho$^1$, 
Joslyn Barnhart$^1$,
Robert Trager$^2$,\\
\textbf{Yoshua Bengio$^3$,}
\textbf{Miles Brundage$^4$,}
\textbf{Allison Carnegie$^5$,}
\textbf{Rumman Chowdhury$^6$,}\\
\textbf{Allan Dafoe$^1$,}
\textbf{Gillian Hadfield$^7$,} 
\textbf{Margaret Levi$^8$,}
\textbf{Duncan Snidal$^9$} \vspace{10pt}\\
 $^1$Google DeepMind, $^2$Blavatnik School of Government, University of Oxford and\\
 Centre for the Governance of AI, $^3$Universit\'{e} de Montr\'{e}al and Mila, CIFAR Fellow,\\
 $^4$OpenAI, 
 $^5$Columbia University,
 $^6$Harvard Berkman Klein,
 $^7$University of Toronto,\\ Vector Institute and OpenAI (independent contractor),
 $^8$Stanford University,
 $^9$Nuffield\\ College, University of Oxford
 }
\begin{document}

\maketitle

\begin{abstract}
International institutions may have an important role to play in
ensuring advanced AI systems benefit humanity. International
collaborations can unlock AI's ability to further sustainable
development, and coordination of regulatory efforts can reduce obstacles
to innovation and the spread of benefits. Conversely, the potential
dangerous capabilities of powerful and general-purpose AI systems
create global externalities in their development and deployment, and
international efforts to further responsible AI practices could help
manage the risks they pose. This paper identifies a set of governance
functions that could be performed at an international level to address
these challenges, ranging from supporting access to frontier AI systems
to setting international safety standards. It groups these functions
into four institutional models that exhibit internal synergies and have
precedents in existing organizations: 1) a Commission on Frontier AI
that facilitates expert consensus on opportunities and risks from
advanced AI, 2) an Advanced AI Governance Organization that sets
international standards to manage global threats from advanced models,
supports their implementation, and possibly monitors compliance with a
future governance regime, 3) a Frontier AI Collaborative that promotes
access to cutting-edge AI, and 4) an AI Safety Project that brings
together leading researchers and engineers to further AI
safety research. We explore the utility of these models and identify
open questions about their viability.
\end{abstract}

\section*{Executive Summary}

Recent advances in AI have highlighted the potentially transformative
impacts of advanced systems.\footnote{By ``advanced AI systems'' we mean systems that are
  highly capable and general purpose.}
  International institutions may have an
important role to play in ensuring these are globally beneficial.

International collaborations could be important for unlocking AI's
abilities to further sustainable development and benefit humanity. Many
societies that could most benefit may not have the resources,
infrastructure or training to take advantage of current cutting-edge AI
systems. Frontier AI development may not focus on global needs, and the
economic benefits of commercial AI technologies could primarily benefit
developed countries. A failure to coordinate or harmonize regulation may
also slow innovation.

Moreover, international efforts may also be necessary to manage the
direct risks posed by powerful AI capabilities. Without adequate
safeguards, some of these capabilities---automated software development,
chemistry and synthetic biology research, text and video
generation---may be misused by malicious actors around the world with
transnational consequences. Furthermore, the significant geopolitical
benefits of rapid AI development decreases the likelihood of adequate AI
governance without international cooperation.

This paper describes a set of international governance functions,
ranging from the distribution of frontier AI systems to the setting of
safety standards, that may be needed to harness the opportunities and
mitigate the risks of advanced AI. Early efforts to perform some of
these functions have been undertaken by inter-governmental organizations
like the Organisation for Economic Co-operation and Development (OECD),
the Global Partnership on AI (GPAI), the Group of 7 (G7) Hiroshima Process,
the International Telecommunication Union (ITU), as well as by private sector 
initiatives like the Partnership on AI, the ML Commons, and International Standards
Organization (ISO) and International Electrotechnical Commission (IEC)
standard-setting initiatives. But the rapid rate of AI progress suggests
further institutional efforts in AI global governance could be needed.

The functions we identify could be divided in multiple ways across
organizations and could involve stakeholders from the public sector, the
private sector, and civil society. We group these functions into four
institutional models that exhibit synergies and have precedents in
existing organizations, and discuss their strengths and limitations:

\begin{itemize}
\item
  \begin{quote}
  An intergovernmental \textbf{Commission on Frontier AI}\footnote{Similar
  institutions include the Intergovernmental Panel on Climate Change (IPCC),
    the Intergovernmental Science-Policy Platform on Biodiversity and
    Ecosystem Services (IPBES), and the Scientific Assessment Panel
    of the United Nations Environment Programme.} could establish a scientific position
  on opportunities and risks from advanced AI and how they may be
  managed. In doing so, it would increase public awareness and
  understanding of AI prospects and issues, contribute to a
  scientifically informed account of AI use and risk mitigation, and be
  a source of expertise for policymakers.
  \end{quote}
\item
  \begin{quote}
  An intergovernmental or multi-stakeholder \textbf{Advanced AI
  Governance Organization}\footnote{Cf.\ the International Civil Aviation
    Organization (ICAO), the International Atomic Energy Agency (IAEA)
    and the Financial Action Task Force (FATF).} could help internationalize
  and align efforts to address global risks from advanced AI systems by
  setting governance norms and standards, and assisting in their
  implementation. It may also perform compliance monitoring functions
  (either independently or in association with industry groups) for an
  international governance regime.
  \end{quote}
\end{itemize}

\begin{itemize}
\item
  \begin{quote}
  A \textbf{Frontier AI Collaborative}\footnote{Cf.\ international
    public-private partnerships like Gavi, the Vaccine Alliance and the
    Global Fund to Fight AIDS, Tuberculosis and Malaria; as well as
    organizations that hold dangerous technologies, like the IAEA's
    nuclear fuel bank and the Atomic Development Authority proposed
    following WWII.} could promote access to advanced AI as an
  international public-private partnership. In doing so, it could help
  underserved societies benefit from cutting-edge AI technology and
  promote international access to AI technology for safety and
  governance objectives.
  \end{quote}
\end{itemize}

\begin{itemize}
\item
  \begin{quote}
  An \textbf{AI Safety Project}\footnote{Cf.\ scientific collaborations
    like the European Organization for Nuclear Research (CERN) and 
    ITER.} could bring together leading researchers and
  engineers, and provide them with access to computing resources and
  advanced AI models for work on technical mitigations of AI risks,
  potentially working with parallel industry efforts. It would promote
  AI safety R\&D by increasing its scale, resourcing and coordination.
  \end{quote}
\end{itemize}

There are important open questions around the viability of such models.
A Commission on Frontier AI will face significant scientific challenges
given the limited scientific research on advanced AI issues, and will
have to combat the politicization of its activities. The rapid rate of
AI progress will make it difficult for an Advanced AI Governance
Organization to set standards that keep up with and are appropriately
scoped for the risk landscape, and the many difficulties of
international coordination raise questions about how participation in an
Organization regime can be incentivized. The potentially dual-use nature
of general purpose AI technologies might hamper a Frontier AI
Collaborative's ability to distribute beneficial systems widely, and the
significant obstacles to underserved societies making use of AI systems
raises questions about its effectiveness as a means of promoting
sustainable development. Finally, an AI Safety Project could struggle to
secure adequate model access to conduct safety research, and it may not
be worthwhile to divert safety researchers away from frontier labs.

Table 1 below summarizes our mapping of institutional functions, the
challenges they address, and the existing and possible institutions that
perform those functions.

There are details about these institutional models that we leave out of
scope or that remain uncertain---whether these institutions should be
new or evolutions of existing organizations, whether the conditions
under which these institutions are likely to be most impactful will
obtain, whether other groupings of institutional functions would be more
effective---but rapid progress on these topics will help prepare for the
development of advanced AI.

\begin{table}[h]
\caption{\label{table}A mapping of international institutional functions, governance objectives and models.}
\vspace{3mm}

\small
\sffamily

\makebox[\textwidth]{
\begin{NiceTabular}{ccccccccc}
\hhline{~--------}

\multicolumn{1}{r|}{Function →} & \multicolumn{4}{Sc|}{\textbf{\begin{tabular}[c]{@{}c@{}}Science and Technology\\ Research, Development and Diffusion\end{tabular}}} & \multicolumn{4}{c|}{\textbf{International Rulemaking and Enforcement}} \\ \hhline{~--------}
\multicolumn{1}{c|}{\begin{tabular}[c]{@{}c@{}}\\~\\Objective /\\institutions ↓\end{tabular}} & \multicolumn{1}{c|}{\begin{tabular}[c]{@{}c@{}}Conduct \\ or Support \\ AI Safety\\ Research\end{tabular}} & \multicolumn{1}{Sc|}{\begin{tabular}[c]{@{}c@{}}Build \\ Consensus on \\ Opportunities \\ and Risks\end{tabular}} & \multicolumn{1}{c|}{\begin{tabular}[c]{@{}c@{}}Develop \\ Frontier AI\end{tabular}} & \multicolumn{1}{c|}{\begin{tabular}[c]{@{}c@{}}Distribute \\ and Enable \\ Access to AI\end{tabular}} & \multicolumn{1}{c|}{\begin{tabular}[c]{@{}c@{}}Set Safety \\ Norms and \\ Standards\end{tabular}} & \multicolumn{1}{c|}{\begin{tabular}[c]{@{}c@{}}Support \\ Implemen-\\ tation of \\ Standards\end{tabular}} & \multicolumn{1}{c|}{\begin{tabular}[c]{@{}c@{}}Monitor \\ Compliance\end{tabular}} & \multicolumn{1}{c|}{\begin{tabular}[c]{@{}c@{}}Control \\ Inputs\end{tabular}} \\ \hline

\multicolumn{1}{|Sc|}{\textbf{\begin{tabular}[c]{@{}c@{}}Spreading\\ Beneficial \\ Technology\end{tabular}}} & \multicolumn{1}{c|}{No} & \multicolumn{1}{c|}{\textcolor{OliveGreen}{Yes}} & \multicolumn{1}{c|}{\textcolor{brown}{Maybe}} & \multicolumn{1}{c|}{\textcolor{OliveGreen}{Yes}} & \multicolumn{1}{c|}{No} & \multicolumn{1}{c|}{No} & \multicolumn{1}{c|}{No} & \multicolumn{1}{c|}{No} \\ \hline

\multicolumn{1}{|Sc|}{\textbf{\begin{tabular}[c]{@{}c@{}}Harmonizing \\ Regulation\end{tabular}}} & \multicolumn{1}{c|}{No} & \multicolumn{1}{c|}{No} & \multicolumn{1}{c|}{No} & \multicolumn{1}{c|}{No} & \multicolumn{1}{c|}{\textcolor{OliveGreen}{Yes}} & \multicolumn{1}{c|}{\textcolor{OliveGreen}{Yes}} & \multicolumn{1}{c|}{No} & \multicolumn{1}{c|}{No} \\ \hline

\multicolumn{1}{|Sc|}{\textbf{\begin{tabular}[c]{@{}c@{}}Ensuring Safe \\ Development \\ and Use\end{tabular}}} & \multicolumn{1}{c|}{\textcolor{brown}{Maybe}} & \multicolumn{1}{c|}{\textcolor{OliveGreen}{Yes}} & \multicolumn{1}{c|}{\textcolor{brown}{Maybe}} & \multicolumn{1}{c|}{\textcolor{brown}{Maybe}} & \multicolumn{1}{c|}{\textcolor{OliveGreen}{Yes}} & \multicolumn{1}{c|}{\textcolor{OliveGreen}{Yes}} & \multicolumn{1}{c|}{\textcolor{brown}{Maybe}} & \multicolumn{1}{c|}{\textcolor{brown}{Maybe}} \\ \hline

\multicolumn{1}{|Sc|}{\textbf{\begin{tabular}[c]{@{}c@{}}Managing \\ Geopolitical \\ Risk Factors\end{tabular}}} & \multicolumn{1}{c|}{No} & \multicolumn{1}{c|}{No} & \multicolumn{1}{c|}{\textcolor{brown}{Maybe}} & \multicolumn{1}{c|}{\textcolor{brown}{Maybe}} & \multicolumn{1}{c|}{No} & \multicolumn{1}{c|}{No} & \multicolumn{1}{c|}{\textcolor{OliveGreen}{Yes}} & \multicolumn{1}{c|}{\textcolor{OliveGreen}{Yes}} \\ \hline
\multicolumn{1}{Sc}{} & \multicolumn{8}{l}{} \\ \hline

\multicolumn{1}{|Sc|}{\textbf{\begin{tabular}[c]{@{}c@{}}Existing Int’l \\ Institutional \\ Efforts\end{tabular}}} & \multicolumn{1}{l|}{} & \multicolumn{1}{c|}{\begin{tabular}[c]{@{}c@{}}OECD, \\ GPAI, G7, \\ ITU\end{tabular}} & \multicolumn{1}{l|}{} & \multicolumn{1}{c|}{} & \multicolumn{1}{c|}{ISO/IEC} & \multicolumn{1}{l|}{} & \multicolumn{1}{l|}{} & \multicolumn{1}{Sc|}{\begin{tabular}[c]{@{}c@{}}Semi-\\ conductor \\ Export\\  Controls\end{tabular}} \\ \hline

\multicolumn{1}{|Sc|}{\textbf{\begin{tabular}[c]{@{}c@{}}Possible \\ Institution\end{tabular}}} & \multicolumn{1}{c|}{\textit{\begin{tabular}[c]{@{}c@{}}AI Safety\\ Project\end{tabular}}} & \multicolumn{1}{c|}{\textit{\begin{tabular}[c]{@{}c@{}}Commission \\ on Frontier AI\end{tabular}}} & \multicolumn{2}{c|}{\textit{Frontier AI Collaborative}} & \multicolumn{3}{c|}{{\color[HTML]{000000} \textit{Advanced AI Governance Agency}}} & \multicolumn{1}{l|}{} \\ \hline

\multicolumn{1}{|c|}{\textbf{{\begin{tabular}[c]{@{}c@{}}Key \\ challenges\end{tabular}}}} & \multicolumn{1}{Sc|}{{\begin{tabular}[c]{@{}c@{}}Model \\ access; \\ diverting \\ talent \end{tabular}}} & \multicolumn{1}{c|}{{\begin{tabular}[c]{@{}c@{}}Politicization; \\ scientific \\ challenges\end{tabular}}} & \multicolumn{2}{c|}{{\begin{tabular}[c]{@{}c@{}}Managing dual-use\\ technology; education, \\ infrastructure and \\ ecosystem obstacles\end{tabular}}} & \multicolumn{3}{c|}{{\begin{tabular}[c]{@{}c@{}}Incentivizing participation; \\ quickly changing risk landscape; \\ maintaining appropriate scope\end{tabular}}} & \multicolumn{1}{l|}{{}} \\ \hline
\end{NiceTabular}
}
\rmfamily

\end{table}

\pagebreak 

\section{Introduction}
The capabilities of AI systems have grown quickly over the last decade.
Employing a growing wealth of algorithmic insights, data sources and
computation power, AI researchers have created systems that can
comprehend language, recognize and generate images and video, write
computer programs and engage in scientific reasoning. If current trends
in AI capabilities continue, AI systems could have transformative
impacts on society.

Powerful AI systems will bring significant benefits and risks. These
systems have the capacity to significantly improve the productivity of
workers and economies as a whole, and to help us address some of our
most important social and technological challenges. But these systems
also present challenges including workforce dislocation, lack of
transparency, biased outcomes, inequitably shared benefits and threats
to national security.

Promoting AI benefits and managing AI risks both have domestic and
international components. On the domestic front, governments and the
private sector will need to establish rules and norms around how
advanced AI systems are developed, distributed, deployed and accessed,
addressing issues like security, distributive impacts, privacy, bias,
and more. A number of challenges have the potential to transcend
national borders and impact societies and economies worldwide.
Accordingly, policymakers, technologists, and AI governance experts have
recently begun to call for specific global AI governance initiatives
centered on international institutions.\footnote{See, e.g., \cite{noauthor_secretary-general_2023, noauthor_elders_2023, dubner_satya_nodate, rees_g20_2023, chowdhury_ai_2023, kakkad_new_2023, hogarth_we_2023}}

This paper contributes to these early conversations by discussing why AI
governance may be needed on an international scale and then offering a
non-exhaustive taxonomy of the institutional functions that
international efforts might require. It explores four possible
international institutions to perform these functions: 1) a Commission
on Frontier AI that facilitates expert consensus on opportunities and
risks from advanced AI; 2) an Advanced AI Governance Organization that
sets international standards, supports their implementation, and could
monitor compliance to future governance regimes; 3) a Frontier AI
Collaborative that develops and distributes cutting-edge AI; and 4) an
AI Safety Project that brings together exceptional researchers,
engineers and compute to further AI safety research. Each of these
approaches seeks to mitigate the societal challenges of advanced AI in
different ways and each confronts significant challenges to its
viability and success.

\section{The Need for International Governance
}

Powerful AI systems have the potential to transform society, economics
and politics in fundamental ways. Because of characteristics like its
high barriers to development/utilization and the possibility of
cross-border use, it is possible that harnessing AI's potential for
global benefit and managing its risks could require governance functions
at the international level.

\subsection*{Promoting Global Benefits}

Access to appropriate AI technology might greatly promote prosperity and
stability \cite{vinuesa_role_2020},
but the benefits might not be evenly distributed or focused on 
the greatest needs of underrepresented communities or
the developing world. Inadequate access to internet services, computing
power, or availability of machine learning training/expertise will also
hinder certain groups' ability to benefit fully from AI advances.

International institutions have long sought to support sustainable
global development. International efforts to \emph{\textbf{build
consensus} \textbf{on AI opportunities}}---especially addressing
barriers to their effective use globally---could support efforts to
\emph{\textbf{distribute and enable access to AI.}} On top of
facilitating access, this could also include building capacity to benefit from AI 
through
education, infrastructure, and local commercial ecosystems. In some
cases, international collaborations (including public-private
partnerships) to \emph{\textbf{develop frontier AI systems}} that are
suited to the needs to underserved communities may also be appropriate.

Inconsistent national regulations could also slow the development and
deployment of AI, as developers of powerful AI technology may be unwilling
to export to countries with inconsistent or unsuitable technology
governance.\footnote{In addition to compliance costs, 
they may be concerned about enabling
  misuse, or the theft of proprietary information \cite{awokuse_stronger_2010}.}
International efforts to \emph{\textbf{set safety norms and standards}}
could help coordinate governance in a way that supports innovation and
serves a broad set of interests.

\subsection*{Managing Shared Risks}

Advanced AI capabilities may also create negative global externalities.
AI systems today are already capable of not just progressing drug
discovery and development, but also of (re)inventing dangerous
chemicals \cite{urbina_dual_2022} and solving foundational problems in synthetic biology \cite{jumper_highly_2021}.
Scientific capabilities like these could be weaponized by malicious
actors for use worldwide.
AI may also be used to create potent cyberweapons that can generate
code, scan codebases for vulnerabilities,
and engineer polymorphic malware in ways that threaten critical
infrastructure \cite{steinhart_what_2023, shimony_chatting_2023}.
Existing AI systems already pose mis- and dis-information issues, and
the introduction of more advanced systems is leading malicious actors to
explore more sophisticated methods of information warfare.\footnote{See, e.g,
  \cite{bond_fake_2023, goldstein_generative_2023}}
Furthermore, building systems that act as intended in novel
circumstances is a challenging problem that may only grow more
difficult \cite{hendrycks_unsolved_2022, amodei_concrete_2016}.
As systems get increasingly capable, there will be greater incentives to
deploy them in higher stakes domains where accidents could have serious
global consequences \cite{arnold_ai_2021, noauthor_statement_2023}.

Implementing protocols for responsible development and deployment will
help address these risks of accident and misuse,\footnote{Potential
  protocols include: training methods that restrict the dangerous
  capabilities and increase the reliability of systems, subjecting AI
  systems to risk assessments that ascertain their propensity to cause
  harm before training or deployment, deployment protocols that secure
  systems against misuse or the exfiltration of its parameters,
  post-deployment monitoring to identify and respond to unforeseen
  risks. See \cite{anderljung_frontier_2023}.} on top of measures
targeted at specific downstream issues.\footnote{Such as the Digital
  Services Act for disinformation, and treaties targeting chemical and
  biological weapons.} However, cross-border access to AI products and
the cross-border effects of misuse and accidents suggests that national
regulation may be ineffective for managing the risks of AI even within
states. States will inevitably be impacted by the development of such
capabilities in other jurisdictions.

To further the international adoption of safety protocols for advanced models, it would be
useful to \emph{\textbf{build consensus on risks}} and how they can be
mitigated\emph{,} and \emph{\textbf{set safety norms and standards}} and
\emph{\textbf{support their implementation}} to help developers and
regulators with responsible development and use. International efforts
to \emph{\textbf{conduct or support AI safety research}} may be
beneficial, if it can increase the rate of safety progress or the reach
of its outputs.

In the longer term, continued algorithmic and hardware progress could
make systems capable of causing significant harm accessible to a large
number of actors, greatly increasing the governance
challenges.\footnote{According to \cite{erdil_algorithmic_2023}
  and \cite{hobbhahn_trends_2022},
  the compute costs of training a model of a fixed performance level
  decreases approximately tenfold every 2 years.} In this case, the
international community might explore measures like
\emph{\textbf{controlling AI inputs}} (although the dual-use/general
purpose nature of the technology creates significant tradeoffs to doing
so) and \emph{\textbf{developing and/or enabling safe forms of access to
AI}}.

The significant geopolitical benefits of AI development may disincline
states to adequately regulate AI: arguments about national
competitiveness are already raised against AI regulation,\footnote{See, e.g., the arguments discussed in \cite{toner_illusion_2023}}
and such pressures may strengthen alongside AI progress. We may
eventually need international agreements that address these geopolitical
risks, with institutions that can \emph{\textbf{monitor compliance}}
where feasible.\footnote{Monitoring can vary significantly in
  intrusiveness and effectiveness: while it will be highly difficult to
  implement adequate monitoring across major geopolitical divides, the
  safety of advanced systems could be a shared interest of major powers
  and a regime to address risk factors from smaller-scale geopolitical
  competition may be feasible.} Efforts to control AI inputs may be
useful to enable non-proliferation of potentially dangerous capabilities
and increase the technical feasibility of monitoring. More
speculatively, efforts to develop frontier AI collectively or distribute
and enable access and its benefits could incentivize participation in a
governance regime.

\begin{center}
    ---
\end{center}

The institutional functions identified above can be summarized and
grouped into the following two broad categories.

\subsubsection*{I.\quad Science and Technology Research, Development and Diffusion}

\begin{itemize}
\item
  \begin{quote}
  \textbf{Conduct or support AI safety research:} 
  Research and develop of measures to reduce the risks
  of AI misuse and accidents stemming from system characteristics like
  dangerous capabilities and unreliability/misalignment. This includes
  work on understanding and evaluating these characteristics and the
  threats they pose, training methods to reduce and manage risky
  behaviors, and examining safe deployment protocols appropriate to
  different system \cite{amodei_concrete_2016, hendrycks_unsolved_2022}.
  \end{quote}
\item
  \begin{quote}
  \textbf{Build consensus on opportunities and risks:} Further
  international understanding of the opportunities and challenges
  created by advanced AI and possible strategies for mitigating the
  risks.
  \end{quote}
\item
  \begin{quote}
  \textbf{Develop frontier AI:} Build cutting-edge AI systems.
  \end{quote}
\item
  \begin{quote}
  \textbf{Distribute and enable access to cutting edge AI:} Facilitate
  access to cutting-edge systems and increase absorptive capacity
  through education, infrastructure, and support of the local commercial
  ecosystem.
  \end{quote}
\end{itemize}

\subsubsection*{II.\quad International Rulemaking and Enforcement}

\begin{itemize}
\item
  \begin{quote}
  \textbf{Set safety norms and standards:} Establish guidelines and
  standards around how AI can be developed, deployed and regulated to
  maximize benefit and minimize risks.
  \end{quote}
\item
  \begin{quote}
  \textbf{Support implementation of standards:} Provide assistance for
  the implementation of established guidelines and standards.
  \end{quote}
\item
  \begin{quote}
  \textbf{Monitor compliance:} Conduct audits /evaluations and issue
  certifications / licenses to ensure adherence to international
  standards and agreements.
  \end{quote}
\item
  \begin{quote}
  \textbf{Control AI inputs:} Manage or monitor models, compute, data
  and other ingredients of potentially dangerous technologies.
  \end{quote}
\end{itemize}

International bodies already perform some of these functions.\footnote{See
  \cite{veale_ai_2023}
  for a more thorough overview.} The OECD's AI Principles and AI Policy
Observatory work, the ITU's AI for Good initiative, and expert reports
from the Global Partnership on AI's Working Group on Responsible AI are
early efforts at building consensus on AI opportunities and risks. Relatedly, the
UK's proposed Foundation Model Taskforce \cite{noauthor_pm_2023}
and the US's proposed Multilateral AI Research Institute could emerge as
multilateral efforts to conduct AI safety research, or
potentially even develop frontier AI systems, though both are in
exploratory phases.\footnote{The amount of funding required to stay on
  the cutting-edge of AI capabilities is significant. See, e.g., \cite{singh_anthropics_2023}.}

Alongside lawmaking efforts like the EU's AI Act and the Council of
Europe's Convention on AI, Human Rights and Democracy, we have seen
early norm and standard setting efforts from ISO/IEC, but little
in the way of implementation support, oversight or certification. In
terms of controlling dangerous inputs: computing resources have been
targeted by US, Japanese and Dutch export controls that prevent the sale
of certain AI chips and semiconductor manufacturing equipment to
China \cite{noauthor_netherlands_2023}.

\section{International Institutions for Advanced
AI}

We have outlined several AI governance functions that might be needed at
an international level, and shown that only a limited number of these
are currently being performed by existing institutions. In this section,
we discuss how functional gaps may be filled.

The functions could be split in multiple ways across institutions:
drawing on existing international organizations and proposals, we
describe four idealized models. We note that the models described in
this section describe roles that could be filled by existing or new
institutions. Participants in these institutions could include
governments, non-governmental organizations, the private sector, and
academia. Table 1 summarizes the previous analysis and the functions of
the institutions we discuss.

\begin{table}[h]
\caption*{Table 1 (repeated): A mapping of international institutional functions, governance objectives and models.}
\vspace{3mm}
\small
\sffamily

\makebox[\textwidth]{
\begin{NiceTabular}{ccccccccc}
\hhline{~--------}

\multicolumn{1}{r|}{Function →} & \multicolumn{4}{Sc|}{\textbf{\begin{tabular}[c]{@{}c@{}}Science and Technology\\ Research, Development and Diffusion\end{tabular}}} & \multicolumn{4}{c|}{\textbf{International Rulemaking and Enforcement}} \\ \hhline{~--------}
\multicolumn{1}{c|}{\begin{tabular}[c]{@{}c@{}}\\~\\Objective /\\institutions ↓\end{tabular}} & \multicolumn{1}{c|}{\begin{tabular}[c]{@{}c@{}}Conduct \\ or Support \\ AI Safety\\ Research\end{tabular}} & \multicolumn{1}{Sc|}{\begin{tabular}[c]{@{}c@{}}Build \\ Consensus on \\ Opportunities \\ and Risks\end{tabular}} & \multicolumn{1}{c|}{\begin{tabular}[c]{@{}c@{}}Develop \\ Frontier AI\end{tabular}} & \multicolumn{1}{c|}{\begin{tabular}[c]{@{}c@{}}Distribute \\ and Enable \\ Access to AI\end{tabular}} & \multicolumn{1}{c|}{\begin{tabular}[c]{@{}c@{}}Set Safety \\ Norms and \\ Standards\end{tabular}} & \multicolumn{1}{c|}{\begin{tabular}[c]{@{}c@{}}Support \\ Implemen-\\ tation of \\ Standards\end{tabular}} & \multicolumn{1}{c|}{\begin{tabular}[c]{@{}c@{}}Monitor \\ Compliance\end{tabular}} & \multicolumn{1}{c|}{\begin{tabular}[c]{@{}c@{}}Control \\ Inputs\end{tabular}} \\ \hline

\multicolumn{1}{|Sc|}{\textbf{\begin{tabular}[c]{@{}c@{}}Spreading\\ Beneficial \\ Technology\end{tabular}}} & \multicolumn{1}{c|}{No} & \multicolumn{1}{c|}{\textcolor{OliveGreen}{Yes}} & \multicolumn{1}{c|}{\textcolor{brown}{Maybe}} & \multicolumn{1}{c|}{\textcolor{OliveGreen}{Yes}} & \multicolumn{1}{c|}{No} & \multicolumn{1}{c|}{No} & \multicolumn{1}{c|}{No} & \multicolumn{1}{c|}{No} \\ \hline

\multicolumn{1}{|Sc|}{\textbf{\begin{tabular}[c]{@{}c@{}}Harmonizing \\ Regulation\end{tabular}}} & \multicolumn{1}{c|}{No} & \multicolumn{1}{c|}{No} & \multicolumn{1}{c|}{No} & \multicolumn{1}{c|}{No} & \multicolumn{1}{c|}{\textcolor{OliveGreen}{Yes}} & \multicolumn{1}{c|}{\textcolor{OliveGreen}{Yes}} & \multicolumn{1}{c|}{No} & \multicolumn{1}{c|}{No} \\ \hline

\multicolumn{1}{|Sc|}{\textbf{\begin{tabular}[c]{@{}c@{}}Ensuring Safe \\ Development \\ and Use\end{tabular}}} & \multicolumn{1}{c|}{\textcolor{brown}{Maybe}} & \multicolumn{1}{c|}{\textcolor{OliveGreen}{Yes}} & \multicolumn{1}{c|}{\textcolor{brown}{Maybe}} & \multicolumn{1}{c|}{\textcolor{brown}{Maybe}} & \multicolumn{1}{c|}{\textcolor{OliveGreen}{Yes}} & \multicolumn{1}{c|}{\textcolor{OliveGreen}{Yes}} & \multicolumn{1}{c|}{\textcolor{brown}{Maybe}} & \multicolumn{1}{c|}{\textcolor{brown}{Maybe}} \\ \hline

\multicolumn{1}{|Sc|}{\textbf{\begin{tabular}[c]{@{}c@{}}Managing \\ Geopolitical \\ Risk Factors\end{tabular}}} & \multicolumn{1}{c|}{No} & \multicolumn{1}{c|}{No} & \multicolumn{1}{c|}{\textcolor{brown}{Maybe}} & \multicolumn{1}{c|}{\textcolor{brown}{Maybe}} & \multicolumn{1}{c|}{No} & \multicolumn{1}{c|}{No} & \multicolumn{1}{c|}{\textcolor{OliveGreen}{Yes}} & \multicolumn{1}{c|}{\textcolor{OliveGreen}{Yes}} \\ \hline
\multicolumn{1}{Sc}{} & \multicolumn{8}{l}{} \\ \hline

\multicolumn{1}{|Sc|}{\textbf{\begin{tabular}[c]{@{}c@{}}Existing Int’l \\ Institutional \\ Efforts\end{tabular}}} & \multicolumn{1}{l|}{} & \multicolumn{1}{c|}{\begin{tabular}[c]{@{}c@{}}OECD, \\ GPAI, G7, \\ ITU\end{tabular}} & \multicolumn{1}{l|}{} & \multicolumn{1}{c|}{} & \multicolumn{1}{c|}{ISO/IEC} & \multicolumn{1}{l|}{} & \multicolumn{1}{l|}{} & \multicolumn{1}{Sc|}{\begin{tabular}[c]{@{}c@{}}Semi-\\ conductor \\ Export\\  Controls\end{tabular}} \\ \hline

\multicolumn{1}{|Sc|}{\textbf{\begin{tabular}[c]{@{}c@{}}Possible \\ Institution\end{tabular}}} & \multicolumn{1}{c|}{\textit{\begin{tabular}[c]{@{}c@{}}AI Safety\\ Project\end{tabular}}} & \multicolumn{1}{c|}{\textit{\begin{tabular}[c]{@{}c@{}}Commission \\ on Frontier AI\end{tabular}}} & \multicolumn{2}{c|}{\textit{Frontier AI Collaborative}} & \multicolumn{3}{c|}{{\color[HTML]{000000} \textit{Advanced AI Governance Agency}}} & \multicolumn{1}{l|}{} \\ \hline

\multicolumn{1}{|c|}{\textbf{{\begin{tabular}[c]{@{}c@{}}Key \\ challenges\end{tabular}}}} & \multicolumn{1}{Sc|}{{\begin{tabular}[c]{@{}c@{}}Model \\ access; \\ diverting \\ talent \end{tabular}}} & \multicolumn{1}{c|}{{\begin{tabular}[c]{@{}c@{}}Politicization; \\ scientific \\ challenges\end{tabular}}} & \multicolumn{2}{c|}{{\begin{tabular}[c]{@{}c@{}}Managing dual-use\\ technology; education, \\ infrastructure and \\ ecosystem obstacles\end{tabular}}} & \multicolumn{3}{c|}{{\begin{tabular}[c]{@{}c@{}}Incentivizing participation; \\ quickly changing risk landscape; \\ maintaining appropriate scope\end{tabular}}} & \multicolumn{1}{l|}{{}} \\ \hline
\end{NiceTabular}
}
\rmfamily

\end{table}

\subsection{Commission on Frontier AI: Fostering
Scientific Consensus
}

There have been several recent proposals of an intergovernmental body to
develop expert consensus on the challenges and opportunities presented
by advanced AI.\footnote{See, e.g., \cite{rees_g20_2023, mailhe_why_2018}.}
Existing institutions like the Intergovernmental Panel on Climate
Change (IPCC), the Intergovernmental Science-Policy Platform on
Biodiversity and Ecosystem Services (IPBES) and the Scientific Assessment
Panel (SAS), which studies ozone depletion under the Montreal Protocol,
provide possible models for an AI-focused scientific institution. Like
these organizations, the Commission on Frontier AI could facilitate
scientific consensus by convening experts to conduct rigorous and
comprehensive assessments of key AI topics, such as interventions to
unlock AI's potential for sustainable development, the effects of AI
regulation on innovation, the distribution of benefits, and possible
dual-use capabilities from advanced systems and how they ought to be
managed.

\subsubsection*{Functions and
Motivation}

International consensus on the opportunities and risks from advanced AI
has the potential to facilitate effective action addressing them, for
example, by engendering a shared desire for the development and adoption
of effective risk mitigation strategies.

Currently, there is significant disagreement even among experts about
the different opportunities and challenges created by advanced
AI,\footnote{See, for example, the disagreement around whether advanced
  AI could pose an extinction risk: \cite{noauthor_statement_2023, noauthor_editorial_2023}.}
and this lack of consensus may worsen over time as the effects of AI
systems increase in scale and number, hindering collective action on the
scale necessary to ensure that AI is developed for the benefit of all.
Furthermore, there are several challenges from advanced AI that may
require international action \emph{before} risks materialize, and the lack of a widely accepted
account or even mapping of AI development trajectories makes it
difficult to take such preparatory actions. Facilitating consensus among
an internationally representative group of experts could be a promising
first step to expanding our levels of confidence in predicting and responding to
technological trends.

\subsubsection*{Challenges and Risks}

\paragraph{Scientific challenges of understanding risks on the horizon:}
Understanding frontier AI risks and their mitigation is technically
challenging. The nature of future AI capabilities and their impact is
difficult to predict, especially given the fast rate of progress. To
increase chances of success, a Commission should foreground scientific
rigor and the selection of highly competent AI experts who work at the
cutting edge of technological development and who can continually
interpret the ever-changing technological and risk landscape.

Unfortunately, there is a relative lack of existing scientific research
on the risks of advanced AI.\footnote{The recent IPCC assessment, for
  reference, was written by 234 scientists from 66 states and drew on
  14,000 scientific papers.} To address the lack of existing scientific
research, a Commission might undertake activities that draw and
facilitate greater scientific attention, such as organizing conferences
and workshops and publishing research agendas. It may be helpful to
write a foundational ``Conceptual Framework''---following the example of
the IPBES---to create a common language and framework that
allows the integration of disparate strands of existing work and paves
the way for future efforts \cite{diaz_ipbes_2015}.

\paragraph{Politicization:} A Commission on Frontier AI would benefit from,
if not require, a clear buffer between experts charged with developing
consensus narratives around the risks and opportunities of AI and
policymakers acting on the political and economic interests of their
states, which might push policies in different directions. 
The scientific understanding of the impacts of AI should ideally 
be seen as a universal good and not be politicized. 

Membership structure can affect a Commission's impartiality and
legitimacy: ideally, there would be broad geographic representation in
the main decisionmaking bodies, and a predominance of scientific experts
in working groups.\footnote{If legitimacy is the primary concern, the
  Commission might adopt the IPCC's innovation of writing key documents
  by consensus, balancing inclusion (states' representatives and
  scientists review, discuss and approve the report line by line) and
  scientific rigor (all suggested amendments must be consisted with
  working group's scientific report that is being summarized) \cite{shaw_relevant_2004}.}
Unfortunately, given the uncertain and controversial nature of advanced
AI risks and opportunities, representation may trade off against a
Commission's ability to overcome scientific challenges and generate
meaningful consensus.\footnote{If it follows the IPCC model, experts
  will be nominated by member states, but there will not be a robust
  climate science discipline to buffer against political interests.} In
addition to striking the correct balance in membership, a Commission
should carefully scope the subject matter of their research---it may,
for example, adopt the IPCC's objective of being ``policy-relevant''
without being ``policy-prescriptive.''

\subsubsection*{Assessment}

The objective of a Commission on Frontier AI is worthwhile in most
circumstances, but the scientific challenges and potential of
politicization imply that a Commission---especially one that aims at
broad political representation---may not be able to build scientific
consensus effectively. The extraordinary pace of technological change
may require more nimble policy responses, such as less institutionalized
and politically authoritative scientific advisory panels on advanced AI.

\subsection{Advanced AI Governance Organization: Promoting Norms and
Standards, Providing Implementation Support, Monitoring
Compliance}

As discussed above, certain misuse and accident risks of advanced AI
systems may pose significant global threats, and international efforts
aimed at managing these risks could be worthwhile. An intergovernmental
or multi-stakeholder organization could perform a variety of governance
functions furthering the regulation of such systems, in particular norm
and standard setting, implementation assistance, and perhaps monitoring
compliance with governance frameworks.\footnote{See, e.g.,
  \cite{noauthor_secretary-general_2023, noauthor_elders_2023,
  altman_governance_2023, chowdhury_ai_2023, marcus_world_2023} for discussions and
  proposals of an institution of this type.}

\subsubsection*{Functions and
Motivation}

We identify two main objectives for an Advanced AI Governance
Organization. How much emphasis it should place on each depends on the
challenges it aims to address.

\paragraph{Objective 1: Internationalizing and harmonizing AI regulation.}
Regulatory regimes that set standards and provide implementation support
may help ensure that powerful AI capabilities do not pose misuse or
accident risks. Standard setting would facilitate widespread
international adoption by: 1) reducing the burden on domestic regulators
to identify necessary safety regulations and protocols, 2) generating
normative pressure for safety protocol adoption, and 3) reducing
frictions around the development of international frameworks.
Implementation support would assist the establishment and maintenance of
regulatory regimes meeting these frameworks. Examples of organizations
that perform similar functions include the Financial Action Task Force
(FATF), the International Telecommunication Union (ITU) and the
International Civil Aviation Organization (ICAO).

The same functions are useful for harmonizing regulation: international
standard setting would reduce cross-border frictions due to differing
domestic regulatory regimes. (It is possible that future regulations
will limit access to powerful AI technologies in jurisdictions with
inadequate AI governance.) Implementation support would help reduce
obstacles to countries meeting international standards and therefore
enable greater access to advanced AI.

\paragraph{Objective 2: Monitoring compliance.} Where states have
incentives to undercut each other\textquotesingle s regulatory
commitments, international institutions may be needed to support and
incentivize best practices. That may require monitoring standards
compliance. At the least intrusive end of the spectrum is self-reporting
of compliance with international standards (as in the Paris
Agreement---see proposals for self-reporting/registration of training
runs \cite{anderljung_frontier_2023}). Organizations like the FATF,
ICAO, and the International Maritime Organization (IMO) take a somewhat
more intrusive approach, monitoring jurisdictions to ensure they adopt
best practice regulations, and in some cases checking on the enforcement
of domestic regulations embodying international standard. In the case of
advanced AI, some observers have asked whether more intrusive forms of
international oversight might be necessary, including detection and
inspections of large data centers (partly analogous to IAEA safeguards).
The more intense and intrusive any monitoring, the more challenging it
may be to get to consensus \cite{shavit_what_2023, toivanen_significance_2017}.

\subsubsection*{Challenges and Risks}

\paragraph{Speed and comprehensiveness in standard setting:} One challenge
for a Governance Organization is that standard setting (especially in an
international and multistakeholder context) tends to be a slow process,
while the rapid and unpredictable nature of frontier AI progress may
require more rapid international action. A Governance Organization may
need to partner with faster-moving expert bodies and expedited
standard-setting approaches. The breadth of membership may also
represent a trade-off between speed and diversity of perspectives.
Broader membership may be important where long-term consensus is
important, while urgent risks may need to be addressed at first by
smaller groups of frontier AI states, or aligned states with relevant
expertise.

\paragraph{Incentivizing participation:} The impact of a Governance
Organization depends on states adopting its standards and/or agreeing to
monitoring. Broad agreement (or agreement among frontier AI states at
least) about the risks that standards and monitoring address and
financial and technical support for standards' implementation may help
induce states' participation. Many states---even those that are not full
members of the organization---adopt FATF standards because they view
them as in their own interests \cite{noauthor_memorandum_2008}.
Other AI-specific incentives for
participation include conditioning on participation access to AI
technology (possibly from a Frontier AI Collaborative) or computing
resources.\footnote{The cloud compute industry and the underlying
  semiconductor supply chain are concentrated in a small number of
  countries.} States might also adopt import restrictions on AI from
countries that are not certified by a Governance Organization---similar,
for instance, to the way states prohibit flights from jurisdictions
without ICAO-certification from entering their airspace.

In the more distant case of high stakes agreements governing AI
development by states (such as arms control treaties), some states may
be especially reluctant to join due to fear of clandestine noncompliance
by other states. They may also worry that international inspections
could compromise state secrets to the benefit of adversaries (which
information security protocols could address in part). Again, the
current reliance of advanced AI development on significant computing
resources may make it easier to track significant AI efforts.\footnote{Oversight 
of data centers may allow the detection
  of large training runs that are subject to international controls. See \cite{shavit_what_2023, brundage_computing_nodate}.} Automated (even
AI-enabled) monitoring may allow closer inspection of large training
runs without compromising secrets. Such measures would likely hinge on
negotiated verification regimes rather than national technical
means---and negotiating verification is always fraught (e.g., in the
case of the Chemical Weapons Convention) and often unsuccessful (e.g.,
in the case of the Biological Weapons Convention)\cite{gallagher_politics_1999, jordan_international_nodate}.

\paragraph{Scoping challenges:} Unlike many other technologies---from
nuclear resources to aviation---AI is already broadly deployed and used
by billions of people every day. To operate efficiently and at
appropriate scale, a Governance Organization should focus primarily on
advanced AI systems that pose significant global risks, but it will be
difficult in practice to decide on the nature and sophistication of AI
tools that should be broadly available and uncontrolled versus the set
of systems that should be subject to national or
international governance. The rapid evolution of these technologies
compounds the problem, as the technological frontier is advancing
quickly, and models that were ``frontier'' a year ago are now both
outdated and widely available.

\subsubsection*{Assessment}

If advanced AI poses misuse and accident risks of a global scope, and
unilateral technical defenses are not sufficient to protect against
them, an international Governance Organization may be valuable. However,
its effectiveness will depend on its membership, governance and
standard-setting processes.

It may be important for governance to apply to all countries, and
particularly to those whose firms are on the frontier of AI development.
Yet, aligned countries may seek to form governance clubs, as they have
in other domains. This facilitates decision-making, but may make it
harder to enlist other countries later in the process. It is unclear
what institutional processes would satisfy the demands of legitimacy and
effectiveness, and incentivize the participation of important groups of
stakeholders.

\subsection{Frontier AI Collaborative: Enabling
International Access to AI
}

Policymakers and pundits have also proposed collaborations to develop
and distribute cutting-edge AI systems, or to ensure such technologies
are accessible to a broad international coalition \cite{hogarth_we_2023, kakkad_new_2023, spirling_why_2023, schmidt_nscai_2021}. Given the significant cost of developing advanced AI systems, a Frontier
AI Collaborative could take the form of an international private-public
partnership that leverages existing technology and capacity in industry,
for example by contracting access to or funding innovation in
appropriate AI technology from frontier AI developers. Such an
organization could draw inspiration from international public-private
partnerships like Gavi - the Vaccine Alliance or The Global Fund to
Fight AIDS, Tuberculosis and Malaria; as well as international
organizations that hold and control powerful technologies, like the
IAEA's nuclear fuel bank \cite{rauf_atoms_2015}
or the Atomic Development Authority that was proposed following
WWII \cite{zaidi_international_2019}.

\subsubsection*{Functions and
Motivation}

A Frontier AI Collaborative could be designed to spread beneficial
technology or serve as a channel for legitimate international access to
advanced AI.

\paragraph{Spreading beneficial technology:} A Collaborative could be
established to ensure the benefits of cutting-edge AI reach groups that
are otherwise underserved by AI development. One motivation for this
objective is that the resources required to develop advanced systems
make their development unavailable to many societies. This may result in
technologies being inadequately designed for and supplied to groups that
may benefit most from them for a variety of reasons:

\begin{enumerate}
\def\labelenumi{\arabic{enumi}.}
\item
  \begin{quote}
  Systems developed by private actors may not adequately cater to all
  societies or demographics: they may not reflect the right values, have
  the right language capabilities, or work efficiently in diverse
  geographies\cite{mohamed_decolonial_2020}.
  \end{quote}
\item
  \begin{quote}
  Private firms may not price their products in ways that allow for
  equitable or broad distribution of benefits.
  \end{quote}
\item
  \begin{quote}
  In order to protect proprietary information, private AI firms may not
  grant deep access to their models (e.g.\ they may restrict API access
  to prevent model imitation \cite{taori_alpaca_2023}),
  which could preclude the development of use cases with significant
  social benefit.
  \end{quote}
\end{enumerate}

A Collaborative could acquire or develop and then distribute AI systems
to address these gaps, pooling resources from member states and
international development programs, working with frontier AI labs to
provide appropriate technology, and partnering with local businesses,
NGOs, and beneficiary governments to better understand technological
needs and overcome barriers to use.\footnote{For example, Gavi promotes
  immunization e.g.\ by funding innovation, and negotiating bulk contracts with
  pharmaceutical companies (especially advanced market commitments) for
  vaccination programs in low-income countries \cite{noauthor_what_2020}.}
It could enable the development of technology that better caters to the
underserved, \cite{mohamed_decolonial_2020} price access to AI models in
a way that is equitable, provide education and build infrastructure to
allow the effective utilization of AI technology, and set a paradigm for
responsible and inclusive AI development. By pooling the resources of
multiple parties towards these ends (including safety talent, which is
currently very scarce in the AI community), one or more of the aims
could potentially be pursued more quickly and effectively than under the
status quo.

\paragraph{Facilitating legitimate international access to powerful AI:}
More speculatively, a sufficiently ambitious, responsible and
legitimately governed AI Collaborative could further AI governance
objectives and reduce geopolitical instability amidst fierce AI
competition among states. For example, membership in a Collaborative and
access to its safe technology could be offered as an incentive for
countries to participate in a governance regime that enforces
responsibility (such as agreements to enact stricter regulation, or
restrict military AI development). The existence of a technologically
empowered neutral coalition may also mitigate the destabilizing effects
of an AI race between states, by reducing the strategic consequences of
one party falling behind or moderating the power concentrated among
competing powers.

In addition, the Collaborative's technology could be used to increase
global resilience to misused or misaligned AI systems by giving experts
a head start in studying the kinds of threats likely to be posed by
other AI systems, and by being deployed for ``protective'' purposes such
as fixing security vulnerabilities in critical infrastructure, detecting
and counteracting disinformation campaigns, identifying misuse or
failures of deployed systems, or monitoring compliance with AI
regulations. This would be especially useful in scenarios where sharply
falling training costs (due to algorithmic progress and Moore's law)
means the ability to train dangerous models is widely spread.

\subsubsection*{Challenges and Risks}

\paragraph{Obstacles to benefiting from AI access:} It is likely to be
difficult to meaningfully empower underserved populations with AI
technology, as the obstacles to their benefiting from AI run much deeper
than access alone. Any Collaborative whose primary objective is global
benefit needs to be adequately integrated into the global development
ecosystem and set up with significant capacity or partnerships for
activities beyond AI development such as: understanding the needs of
member countries, building absorptive capacity through education and
infrastructure, and supporting the development of a local commercial
ecosystem to make use of the technology \cite{noauthor_unfccc_nodate, maya_capacity_2010}.
The resources required to overcome these obstacles is likely to be
substantial, and it is unclear whether such a Collaborative would be an
effective means of promoting development.

\paragraph{Diffusion of dual-use technologies:} Another challenge for the
Collaborative would be managing the risk of diffusing dangerous
technologies. On the one hand, in order to fulfill its objectives, the
Collaborative would need to significantly promote access to the benefits
of advanced AI (objective 1), or put control of cutting-edge AI
technology in the hands of a broad coalition (objective 2). On the other
hand, it may be difficult to do this without diffusing dangerous AI
technologies around the world, if the most powerful AI systems are
general purpose, dual-use, and proliferate easily.\footnote{For example:
  it may be difficult to protect broadly-deployed models from imitation,
  and it may be difficult to secure the deployment pipeline from
  attempts to copy model weights.} This is especially the
case if the Collaborative aims to deploy cutting-edge general purpose
systems to manage AI risks: the kinds of systems (and their underlying
source code and algorithms) capable of meaningfully protecting against
dangerous AI or furthering governance objectives may pose an exceptional
misuse risk, as they will likely be engineered from highly capable,
general purpose models.

To address such a challenge, it would be important for the Collaborative
to have a clear mandate and purpose. Members of a Collaborative would
need to have a strong understanding of the risks of the models being
developed now and in the future, and their implications for model
distribution, organization security (especially restrictions on the
movement of Collaborative model weights), and other activities that may
impact their ability to benefit from the Collaborative. Only by doing
this would the Collaborative be able to consistently implement the
necessary controls to manage frontier systems. It may also be necessary
to exclude from participation states who are likely to want to use AI
technology in non-peaceful ways, or make participation in a governance
regime the precondition for membership.

\subsubsection*{Assessment}

A Frontier AI Collaborative may indeed be a viable way of spreading AI
benefits. However, the significant obstacles to societies benefiting
from AI access raise questions about its competitiveness (relative to
other development initiatives) as a means of promoting the welfare of
underserved communities.

The viability of a Collaborative as a site of legitimately controlled
technology also unclear: it depends on whether a balance can be struck
between legitimately pursuing technologies powerful enough to positively
affect international stability, and managing the proliferation of
dangerous systems.

\subsection{AI Safety Project: Conducting Technical Safety
Research}

The final model we discuss is an international collaboration to conduct technical AI
safety research\footnote{This could include work on
  understanding and evaluating characteristics of systems such as
  alignment/reliability and dangerous capabilities, training methods to
  reduce and manage these characteristics, and deployment protocols
  (such as system security, monitoring, accident-response) that are
  appropriate to different system characteristics.} at an ambitious
scale.\footnote{See, e.g., \cite{dubner_satya_nodate, hammond_opinion_2023}}

Tthe Safety Project would be modeled after
large-scale scientific collaborations like ITER and CERN. Concretely, it
would be an institution with significant compute, engineering capacity
and access to models (obtained via agreements with leading AI
developers), and would recruit the world's leading experts in AI, AI
safety and other relevant fields to work collaboratively on how to
engineer and deploy advanced AI systems such that they are reliable and
less able to be misused. CERN and ITER are intergovernmental
collaborations; we note that an AI Safety Project need not be, and should be
organized to benefit from the AI Safety expertise in civil society and 
the private sector.

\subsubsection*{Functions and
Motivation}

The motivation behind an international Safety Project would be to
accelerate AI safety research by increasing its scale, resourcing and
coordination, thereby expanding the ways in which AI can be safely
deployed, and mitigating risks stemming from powerful general purpose
capabilities.\footnote{Being a public good, AI safety may be underfunded
  by default, which the Safety Project would address as a site of
  collective contribution.} Researchers---including those who would not
otherwise be working on AI safety---could be drawn by its international
stature and enabled by the project's exceptional compute, engineers and
model access. The Project would become a vibrant research community that
benefits from tighter information flows and a collective focus on AI
safety. The Project should also have exceptional leaders and governance
structures that ensure its efforts are most effectively targeted at
critical questions on the path to safer AI systems.

Because perceptions of AI risk vary around the world, such an effort
would likely be spearheaded by frontier risk-conscious actors like the
US and UK governments, AGI labs and civil society groups. In the long
run, it would be important for membership to be broad to ensure its
research is recognized and informs AI development and deployment around
the world.\footnote{While safety-relevant insights should be publicized
  for international use, other innovations with commercial value can be
  collectively owned by or affordably licensed to member states to
  incentivize broad participation. See, e.g., CERN's approach to this
  \cite{le_gall_how_2021}.}

\subsubsection*{Challenges and Risks}

\paragraph{Pulling safety research away from frontier developers:} One potential
effect of this model is that it diverts safety research away from the
sites of frontier AI development. It is possible that safety research
is best conducted in close proximity to AI development to deepen safety
researchers' understanding of the processes and systems they are trying
to make safe and to ensure there is adequate safety expertise in-house.
This risk could be addressed by offering safety researchers within AI
labs dual appointments or advisory roles in the Project, and may become
less of an issue if the \emph{practice} of AI safety becomes
institutionalized and separated from research and development.

\paragraph{Security concerns and model access:} In order to be effective,
participants in the Project need to have access to advanced AI models,
which may allow them to illegally copy the model's weights, clone the
model via access to its outputs \cite{taori_alpaca_2023},
or understand how it could be replicated (by determining its
architecture or training process). Given the importance of these assets
to the business interests of frontier labs, it may be difficult to
negotiate agreements where adequate model access is granted. It may also
lead to the diffusion of dangerous technologies.

This issue may be addressed by restricting membership in the Safety
Project and by information security measures. In particular, it may be
possible to silo information, structure model access, and design
internal review processes in such a way that meaningfully reduces this
risk while ensuring research results are subject to adequate scientific
scrutiny. Certain types of research, such as the development of model
evaluations and red-teaming protocols, can proceed effectively with API
access to the models, while others such as mechanistic interpretability
will require access to the model weights and architectures, but may not
need to work with the latest (and therefore most sensitive)
models \cite{shevlane_structured_2022}.

\subsubsection*{Assessment}

Technical progress on how to increase the reliability of advanced AI
systems and protect them from misuse will likely be a priority in AI
governance. It remains to be seen whether---due to issues of model
access and the allocation of experts between a Safety Project and sites
of frontier AI development---an AI Safety Project will be the most
effective way of pursuing this goal.

\subsection{Combining Institutional
Functions}

We can imagine institutions taking on the role of several of the models
above. For example, the Commission on Frontier AI and the AI Safety
Project make an obvious pairing: a Commission could scale up research
functions to supplement the synthesis and consensus-building efforts, or
a Project could conduct synthesis work in the course of its activities
and gradually take on a consensus-establishing role. A Frontier AI
Collaborative would also likely conduct safety research, and could
easily absorb additional resourcing to become a world-leading Safety
Project.

\section{Conclusion}

This paper has outlined several reasons why the world may want to expand
existing initiatives in AI governance and safety and discussed the
strengths and limitations of four possible institutional models to
address these needs.

To better harness advanced AI for global benefit, international efforts
to help underserved societies access and use advanced AI systems will be
important. A Frontier AI Collaborative that acquires and distributes AI
systems could be helpful, if it can effectively enable underserved
groups to take full advantage of such systems. A Commission on Frontier
AI could help identify the areas where international efforts can most
effectively achieve these goals, if it can prevent the politicization of
its work. Relatedly, it will be important for governance approaches
around the world to be coordinated, so as to reduce frictions to
innovation and access: an Advanced AI Governance Organization that sets
international standards for governance of the most advanced models could facilitate this.

To manage global risks from powerful AI systems, effective AI governance
regimes may be needed around the world. An Advanced AI Governance
Organization that establishes governance frameworks for managing global
threats from advanced systems and helps with their implementation may
help internationalize effective regulatory measures, but it may be
difficult to establish reliable standards if AI progress continues at
the present rate, and also difficult to incentivize adoption of an
Organization's standards if there is a lack of global consensus on AI
risks. A Commission on Frontier AI could also support governance efforts
by building scientific consensus around AI risks and their mitigation,
although its task of providing a scientifically credible and
internationally recognized account of a quickly changing risk landscape
will be challenging, especially given the relative lack of existing
scientific research on the topic. An AI Safety Project could accelerate
the rate at which technical methods of mitigating AI risks are
developed---provided it can overcome its efficiency and model access
hurdles, and a Frontier AI Collaborative's technology might be used to
increase global resilience to misused or misaligned AI systems. More
speculatively, the functions of a Governance Organization and
Collaborative could in some cases counteract the geopolitical factors
exacerbating AI risks.

The taxonomy of functions we have presented is not exhaustive, nor do we
argue that our institutional grouping is the most promising. Given the
immense global opportunities and challenges presented by AI systems that
may be on the horizon, the topic of international institutions for AI
governance demands much greater analytical and practical attention.

\section*{Acknowledgements}
We are grateful to the following people for discussion and input: Michael Aird, Jeff Alstott, Jon Bateman, Alexandra Belias, Dorothy Chou, Jack Clark, Lukas Finnveden, Iason Gabriel, Ben Garfinkel, Erich Grunewald, Oliver Guest, Jackie Kay, Noam Kolt, Sebastien Krier, Lucy Lim, Nicklas Lundblad, Stewart Patrick, George Perkovich, Toby Shevlane, Kent Walker and Ankur Vora.
We would also like to thank participants of the September 2022 and June 2023 IGOTAI Seminars, in which
early work was discussed.

\bibliographystyle{plainurl}
\bibliography{refs}

\begin{thebibliography}{10}

\bibitem{noauthor_netherlands_2023}
The {Netherlands} joins the {U}.{S}. in restricting semiconductor exports to
  {China}.
\newblock Allen Overy, March 2023.
\newblock URL:
  \url{https://www.allenovery.com/en-gb/global/news-and-insights/publications/the-netherlands-joins-the-us-in-restricting-semiconductor-exports-to-china}.

\bibitem{altman_governance_2023}
Sam Altman, Greg Brockman, and Ilya Sutskever.
\newblock Governance of superintelligence.
\newblock OpenAI, 2023.
\newblock URL: \url{https://openai.com/blog/governance-of-superintelligence}.

\bibitem{amodei_concrete_2016}
Dario Amodei, Chris Olah, Jacob Steinhardt, Paul Christiano, John Schulman, and
  Dan Mané.
\newblock Concrete {Problems} in {AI} {Safety}, July 2016.
\newblock arXiv:1606.06565 [cs].
\newblock URL: \url{http://arxiv.org/abs/1606.06565}.

\bibitem{anderljung_frontier_2023}
Markus Anderljung, Joslyn Barnhart, Jade Leung, Anton Korinek, Cullen O'Keefe,
  Jess Whittlestone, Shahar Avin, Miles Brundage, Justin Bullock, Duncan
  Cass-Beggs, Ben Chang, Tantum Collins, Tim Fist, Gillian Hadfield, Alan
  Hayes, Lewis Ho, Sarah Hooker, Eric Horvitz, Noam Kolt, Jonas Schuett,
  Yonadav Shavit, Divya Siddarth, Robert Trager, and Kevin Wolf.
\newblock Frontier {AI} {Regulation}: {Managing} {Emerging} {Risks} to {Public}
  {Safety}, 2023.

\bibitem{arnold_ai_2021}
Zachary Arnold and Helen Toner.
\newblock {AI} {Accidents}: {An} {Emerging} {Threat}.
\newblock Center for Security and Emerging Technology, 2021.
\newblock URL:
  \url{https://cset.georgetown.edu/publication/ai-accidents-an-emerging-threat/}.

\bibitem{awokuse_stronger_2010}
Titus Awokuse and Hong Yin.
\newblock Do {Stronger} {Intellectual} {Property} {Rights} {Protection}
  {Induce} {More} {Bilateral} {Trade}? {Evidence} from {China}'s {Imports}.
\newblock {\em World Development}, 38(8), 2010.

\bibitem{bond_fake_2023}
Shannon Bond.
\newblock Fake viral images of an explosion at the {Pentagon} were probably
  created by {AI}.
\newblock {\em NPR}, May 2023.
\newblock URL:
  \url{https://www.npr.org/2023/05/22/1177590231/fake-viral-images-of-an-explosion-at-the-pentagon-were-probably-created-by-ai}.

\bibitem{brundage_computing_nodate}
Miles Brundage, Girish Sastry, Lennart Heim, Haydn Belfield, Julian Hazell,
  Markus Anderljung, Shahar Avin, Jade Leung, Cullen O'Keefe, and Richard Ngo.
\newblock Computing {Power} and the {Governance} of {Artificial}
  {Intelligence}.

\bibitem{noauthor_statement_2023}
Statement on {AI} {Risk}.
\newblock Center for AI Safety, 2023.
\newblock URL: \url{https://www.safe.ai/statement-on-ai-risk}.

\bibitem{chowdhury_ai_2023}
Rumman Chowdhury.
\newblock {AI} {Desperately} {Needs} {Global} {Oversight}.
\newblock {\em Wired}, 2023.
\newblock Section: tags.
\newblock URL:
  \url{https://www.wired.com/story/ai-desperately-needs-global-oversight/}.

\bibitem{dubner_satya_nodate}
Stephen Dubner.
\newblock Satya {Nadella}’s {Intelligence} {Is} {Not} {Artificial}.
\newblock URL:
  \url{https://freakonomics.com/podcast/satya-nadellas-intelligence-is-not-artificial/}.

\bibitem{diaz_ipbes_2015}
Sandra Díaz, Sebsebe Demissew, Julia Carabias, Carlos Joly, Mark Lonsdale,
  Neville Ash, Anne Larigauderie, Jay~Ram Adhikari, Salvatore Arico, András
  Báldi, Ann Bartuska, Ivar~Andreas Baste, Adem Bilgin, Eduardo Brondizio,
  Kai~MA Chan, Viviana~Elsa Figueroa, Anantha Duraiappah, Markus Fischer,
  Rosemary Hill, Thomas Koetz, Paul Leadley, Philip Lyver, Georgina~M Mace,
  Berta Martin-Lopez, Michiko Okumura, Diego Pacheco, Unai Pascual,
  Edgar~Selvin Pérez, Belinda Reyers, Eva Roth, Osamu Saito, Robert~John
  Scholes, Nalini Sharma, Heather Tallis, Randolph Thaman, Robert Watson,
  Tetsukazu Yahara, Zakri~Abdul Hamid, Callistus Akosim, Yousef Al-Hafedh,
  Rashad Allahverdiyev, Edward Amankwah, Stanley~T Asah, Zemede Asfaw, Gabor
  Bartus, L~Anathea Brooks, Jorge Caillaux, Gemedo Dalle, Dedy Darnaedi, Amanda
  Driver, Gunay Erpul, Pablo Escobar-Eyzaguirre, Pierre Failler, Ali
  Moustafa~Mokhtar Fouda, Bojie Fu, Haripriya Gundimeda, Shizuka Hashimoto,
  Floyd Homer, Sandra Lavorel, Gabriela Lichtenstein, William~Armand Mala,
  Wadzanayi Mandivenyi, Piotr Matczak, Carmel Mbizvo, Mehrasa Mehrdadi,
  Jean~Paul Metzger, Jean~Bruno Mikissa, Henrik Moller, Harold~A Mooney, Peter
  Mumby, Harini Nagendra, Carsten Nesshover, Alfred~Apau Oteng-Yeboah, György
  Pataki, Marie Roué, Jennifer Rubis, Maria Schultz, Peggy Smith, Rashid
  Sumaila, Kazuhiko Takeuchi, Spencer Thomas, Madhu Verma, Youn Yeo-Chang, and
  Diana Zlatanova.
\newblock The {IPBES} {Conceptual} {Framework} — connecting nature and
  people.
\newblock {\em Current Opinion in Environmental Sustainability}, 14:1--16, June
  2015.
\newblock URL:
  \url{https://www.sciencedirect.com/science/article/pii/S187734351400116X},
  \href {https://doi.org/10.1016/j.cosust.2014.11.002}
  {\path{doi:10.1016/j.cosust.2014.11.002}}.

\bibitem{noauthor_elders_2023}
The {Elders} urge global co-operation to manage risks and share benefits of
  {AI}.
\newblock The Elders, May 2023.
\newblock URL:
  \url{https://theelders.org/news/elders-urge-global-co-operation-manage-risks-and-share-benefits-ai}.

\bibitem{erdil_algorithmic_2023}
Ege Erdil and Tamay Besiroglu.
\newblock Algorithmic {Progress} in {Computer} {Vision}, 2023.
\newblock URL: \url{https://arxiv.org/abs/2212.05153}.

\bibitem{gallagher_politics_1999}
Nancy Gallagher.
\newblock {\em The {Politics} of {Verification}}.
\newblock Johns Hopkins University Press, 1999.

\bibitem{noauthor_what_2020}
What is an {Advance} {Market} {Commitment} and how could it help beat
  {COVID}-19? {\textbar} {Gavi}, the {Vaccine} {Alliance}.
\newblock Gavi, the Vaccine Alliance, 2020.
\newblock URL:
  \url{https://www.gavi.org/vaccineswork/what-advance-market-commitment-and-how-could-it-help-beat-covid-19}.

\bibitem{goldstein_generative_2023}
Josh~A. Goldstein, Girish Sastry, Micah Musser, Renee DiResta, Matthew Gentzel,
  and Katerina Sedova.
\newblock Generative {Language} {Models} and {Automated} {Influence}
  {Operations}: {Emerging} {Threats} and {Potential} {Mitigations}, January
  2023.
\newblock arXiv:2301.04246 [cs].
\newblock URL: \url{http://arxiv.org/abs/2301.04246}, \href
  {https://doi.org/10.48550/arXiv.2301.04246}
  {\path{doi:10.48550/arXiv.2301.04246}}.

\bibitem{noauthor_pm_2023}
{PM} urges tech leaders to grasp generational opportunities and challenges of
  {AI}.
\newblock GOV.UK, 2023.
\newblock URL:
  \url{https://www.gov.uk/government/news/pm-urges-tech-leaders-to-grasp-generational-opportunities-and-challenges-of-ai}.

\bibitem{hammond_opinion_2023}
Samuel Hammond.
\newblock Opinion {\textbar} {We} {Need} a {Manhattan} {Project} for {AI}
  {Safety}.
\newblock POLITICO, May 2023.
\newblock URL:
  \url{https://www.politico.com/news/magazine/2023/05/08/manhattan-project-for-ai-safety-00095779}.

\bibitem{hendrycks_unsolved_2022}
Dan Hendrycks, Nicholas Carlini, John Schulman, and Jacob Steinhardt.
\newblock Unsolved {Problems} in {ML} {Safety}, June 2022.
\newblock arXiv:2109.13916 [cs].
\newblock URL: \url{http://arxiv.org/abs/2109.13916}.

\bibitem{hobbhahn_trends_2022}
Marius Hobbhahn.
\newblock Trends in {GPU} price-performance.
\newblock Epoch, 2022.
\newblock URL: \url{https://epochai.org/blog/trends-in-gpu-price-performance}.

\bibitem{hogarth_we_2023}
Ian Hogarth.
\newblock We must slow down the race to {God}-like {AI} {\textbar} {Financial}
  {Times}, 2023.
\newblock URL:
  \url{https://www.ft.com/content/03895dc4-a3b7-481e-95cc-336a524f2ac2}.

\bibitem{jordan_international_nodate}
Richard Jordan, Nicholas Emery-Xu, and Robert Trager.
\newblock International {Governance} of {Advanced} {AI}.

\bibitem{jumper_highly_2021}
John Jumper, Richard Evans, Alexander Pritzel, Tim Green, Michael Figurnov,
  Olaf Ronneberger, Kathryn Tunyasuvunakool, Russ Bates, Augustin Žídek, Anna
  Potapenko, Alex Bridgland, Clemens Meyer, Simon A.~A. Kohl, Andrew~J.
  Ballard, Andrew Cowie, Bernardino Romera-Paredes, Stanislav Nikolov, Rishub
  Jain, Jonas Adler, Trevor Back, Stig Petersen, David Reiman, Ellen Clancy,
  Michal Zielinski, Martin Steinegger, Michalina Pacholska, Tamas Berghammer,
  Sebastian Bodenstein, David Silver, Oriol Vinyals, Andrew~W. Senior, Koray
  Kavukcuoglu, Pushmeet Kohli, and Demis Hassabis.
\newblock Highly accurate protein structure prediction with {AlphaFold}.
\newblock {\em Nature}, 596(7873):583--589, August 2021.
\newblock Number: 7873 Publisher: Nature Publishing Group.
\newblock URL: \url{https://www.nature.com/articles/s41586-021-03819-2}, \href
  {https://doi.org/10.1038/s41586-021-03819-2}
  {\path{doi:10.1038/s41586-021-03819-2}}.

\bibitem{kakkad_new_2023}
Jeegar Kakkad, Benedict Macon-Cooney, Jess Northend, James Phillips, Nitarshan
  Rajkumar, Luke Stanley, and Tom Westgarth.
\newblock A {New} {National} {Purpose}: {Innovation} {Can} {Power} the {Future}
  of {Britain}, 2023.
\newblock URL:
  \url{https://www.institute.global/insights/politics-and-governance/new-national-purpose-innovation-can-power-future-britain}.

\bibitem{le_gall_how_2021}
Antoine Le~Gall.
\newblock How {CERN} intellectual property helps entrepreneurship.
\newblock CERN, 2021.
\newblock URL:
  \url{https://home.cern/news/news/knowledge-sharing/how-cern-intellectual-property-helps-entrepreneurship}.

\bibitem{mailhe_why_2018}
Nicolas Mailhe.
\newblock Why {We} {Need} an {Intergovernmental} {Panel} for {Artificial}
  {Intelligence} - {Our} {World}, 2018.
\newblock URL:
  \url{https://ourworld.unu.edu/en/why-we-need-an-intergovernmental-panel-for-artificial-intelligence}.

\bibitem{marcus_world_2023}
Gary Marcus and Anka Reuel.
\newblock The world needs an international agency for artificial intelligence,
  say two {AI} experts.
\newblock {\em The Economist}, 2023.
\newblock URL:
  \url{https://www.economist.com/by-invitation/2023/04/18/the-world-needs-an-international-agency-for-artificial-intelligence-say-two-ai-experts}.

\bibitem{maya_capacity_2010}
Shakespeare Maya.
\newblock Capacity {Building} for {Technology} {Transfer} in the {African}
  {Context}: {Priorities} and {Strategies}.
\newblock {\em UNFCC}, 2010.

\bibitem{mohamed_decolonial_2020}
Shakir Mohamed, Marie-Therese Png, and William Isaac.
\newblock Decolonial {AI}: {Decolonial} {Theory} as {Sociotechnical}
  {Foresight} in {Artificial} {Intelligence}.
\newblock {\em Philosophy \& Technology}, 33(4):659--684, December 2020.
\newblock \href {https://doi.org/10.1007/s13347-020-00405-8}
  {\path{doi:10.1007/s13347-020-00405-8}}.

\bibitem{noauthor_editorial_2023}
Editorial: {Stop} talking about tomorrow’s {AI} doomsday when {AI} poses
  risks today.
\newblock {\em Nature}, 618(7967):885--886, June 2023.
\newblock Bandiera\_abtest: a Cg\_type: Editorial Number: 7967 Publisher:
  Nature Publishing Group Subject\_term: Machine learning, Authorship, Ethics.
\newblock URL: \url{https://www.nature.com/articles/d41586-023-02094-7}.

\bibitem{rauf_atoms_2015}
Tariq Rauf.
\newblock From ‘{Atoms} for {Peace}’ to an {IAEA} {Nuclear} {Fuel} {Bank}
  {\textbar} {Arms} {Control} {Association}.
\newblock Arms Control Association, October 2015.
\newblock URL:
  \url{https://www.armscontrol.org/act/2015-10/features/%E2%80%98atoms-peace%E2%80%99-iaea-nuclear-fuel-bank}.

\bibitem{rees_g20_2023}
Martin Rees, Shivaji Sondhi, and Krishnaswamy VijayRaghavan.
\newblock G20 must set up an international panel on technological change.
\newblock Hindustan Times, March 2023.
\newblock URL:
  \url{https://www.hindustantimes.com/opinion/g20-must-set-up-an-international-panel-on-technological-change-101679237287848.html}.

\bibitem{schmidt_nscai_2021}
Eric Schmidt, Robert Work, Safra Catz, Eric Horvitz, Steve Chien, Andrew Jassy,
  Mignon Clyburn, Gilman Louie, Chris Darby, William Mark, Kenneth Ford, Jason
  Matheny, Jose-Marie Griffiths, Katharina McFarland, and Andrew Moore.
\newblock {NSCAI} {Final} {Report}.
\newblock Technical report, National Security Commission on Artificial
  Intelligence, 2021.
\newblock URL: \url{https://www.nscai.gov/2021-final-report/}.

\bibitem{shavit_what_2023}
Yonadav Shavit.
\newblock What does it take to catch a {Chinchilla}? {Verifying} {Rules} on
  {Large}-{Scale} {Neural} {Network} {Training} via {Compute} {Monitoring}, May
  2023.
\newblock arXiv:2303.11341 [cs].
\newblock URL: \url{http://arxiv.org/abs/2303.11341}, \href
  {https://doi.org/10.48550/arXiv.2303.11341}
  {\path{doi:10.48550/arXiv.2303.11341}}.

\bibitem{shaw_relevant_2004}
Alison Shaw and John Robinson.
\newblock Relevant {But} {Not} {Prescriptive}: {Science} {Policy} {Models}
  within the {IPCC}.
\newblock {\em Philosophy Today}, 48:84--95, January 2004.
\newblock \href {https://doi.org/10.5840/philtoday200448Supplement9}
  {\path{doi:10.5840/philtoday200448Supplement9}}.

\bibitem{shevlane_structured_2022}
Toby Shevlane.
\newblock Structured access: an emerging paradigm for safe {AI} deployment,
  April 2022.
\newblock arXiv:2201.05159 [cs].
\newblock URL: \url{http://arxiv.org/abs/2201.05159}, \href
  {https://doi.org/10.48550/arXiv.2201.05159}
  {\path{doi:10.48550/arXiv.2201.05159}}.

\bibitem{shimony_chatting_2023}
Eran Shimony and Omar Tsarfati.
\newblock Chatting {Our} {Way} {Into} {Creating} a {Polymorphic} {Malware},
  2023.
\newblock URL:
  \url{https://www.cyberark.com/resources/threat-research-blog/chatting-our-way-into-creating-a-polymorphic-malware}.

\bibitem{singh_anthropics_2023}
Kyle~Wiggers Singh, Devin Coldewey {and}~Manish.
\newblock Anthropic's \${5B}, 4-year plan to take on {OpenAI}.
\newblock TechCrunch, April 2023.
\newblock URL:
  \url{https://techcrunch.com/2023/04/06/anthropics-5b-4-year-plan-to-take-on-openai/}.

\bibitem{spirling_why_2023}
Arthur Spirling.
\newblock Why open-source generative {AI} models are an ethical way forward for
  science.
\newblock {\em Nature}, 616(7957):413--413, April 2023.
\newblock Bandiera\_abtest: a Cg\_type: World View Number: 7957 Publisher:
  Nature Publishing Group Subject\_term: Ethics, Machine learning, Technology,
  Scientific community.
\newblock URL: \url{https://www.nature.com/articles/d41586-023-01295-4}, \href
  {https://doi.org/10.1038/d41586-023-01295-4}
  {\path{doi:10.1038/d41586-023-01295-4}}.

\bibitem{steinhart_what_2023}
Jacob Steinhart.
\newblock What will {GPT}-2030 look like?
\newblock Bounded Regret, June 2023.
\newblock URL:
  \url{https://bounded-regret.ghost.io/what-will-gpt-2030-look-like/}.

\bibitem{taori_alpaca_2023}
Rohan Taori, Ishaan Gulrajani, Tianyi Zhang, Yann Dubois, Xuechen Li, Carlos
  Guestrin, Percy Liang, and Tatsunori Hashimoto.
\newblock Alpaca: {A} {Strong}, {Replicable} {Instruction}-{Following} {Model}.
\newblock Stanford CRFM, March 2023.
\newblock URL: \url{https://crfm.stanford.edu/2023/03/13/alpaca.html}.

\bibitem{toivanen_significance_2017}
Henrietta Toivanen.
\newblock The {Significance} of {Strategic} {Foresight} in {Verification}
  {Technologies}: {A} {Case} {Study} of the {INF} {Treaty}.
\newblock Technical Report LLNL-TR--738786, 1502006, 892173, September 2017.
\newblock URL: \url{https://www.osti.gov/servlets/purl/1502006/}, \href
  {https://doi.org/10.2172/1502006} {\path{doi:10.2172/1502006}}.

\bibitem{toner_illusion_2023}
Helen Toner, Jenny Xiao, and Jeffrey Ding.
\newblock The {Illusion} of {China}’s {AI} {Prowess}.
\newblock {\em Foreign Affairs}, June 2023.
\newblock URL:
  \url{https://www.foreignaffairs.com/china/illusion-chinas-ai-prowess-regulation}.

\bibitem{noauthor_memorandum_2008}
Memorandum by the {Financial} {Action} {Task} {Force} ({FATF}) {Secretariat}.
\newblock UK Parliament, 2008.
\newblock URL:
  \url{https://publications.parliament.uk/pa/ld200809/ldselect/ldeucom/132/132we08.htm}.

\bibitem{noauthor_secretary-general_2023}
Secretary-{General} {Urges} {Broad} {Engagement} from {All} {Stakeholders}
  towards {United} {Nations} {Code} of {Conduct} for {Information} {Integrity}
  on {Digital} {Platforms} {\textbar} {UN} {Press}.
\newblock United Nations, June 2023.
\newblock URL: \url{https://press.un.org/en/2023/sgsm21832.doc.htm}.

\bibitem{noauthor_unfccc_nodate}
{UNFCCC} {Technology} {Mechanism}.
\newblock UNFCCC.
\newblock URL:
  \url{https://unfccc.int/ttclear/support/technology-mechanism.html}.

\bibitem{urbina_dual_2022}
Fabio Urbina, Filippa Lentzos, Cédric Invernizzi, and Sean Ekins.
\newblock Dual use of artificial-intelligence-powered drug discovery.
\newblock {\em Nature Machine Intelligence}, 4(3):189--191, March 2022.
\newblock Number: 3 Publisher: Nature Publishing Group.
\newblock URL: \url{https://www.nature.com/articles/s42256-022-00465-9}, \href
  {https://doi.org/10.1038/s42256-022-00465-9}
  {\path{doi:10.1038/s42256-022-00465-9}}.

\bibitem{veale_ai_2023}
Michael Veale, Kira Matus, and Robert Gorwa.
\newblock {AI} and {Global} {Governance}: {Modalities}, {Rationales},
  {Tensions}.
\newblock {\em Annual Review of Law and Social Science}, 19, 2023.

\bibitem{vinuesa_role_2020}
Ricardo Vinuesa, Hossein Azizpour, Iolanda Leite, Madeline Balaam, Virginia
  Dignum, Sami Domisch, Anna Felländer, Simone~Daniela Langhans, Max Tegmark,
  and Francesco Fuso~Nerini.
\newblock The role of artificial intelligence in achieving the {Sustainable}
  {Development} {Goals}.
\newblock {\em Nature Communications}, 11(1):233, January 2020.
\newblock Number: 1 Publisher: Nature Publishing Group.
\newblock URL: \url{https://www.nature.com/articles/s41467-019-14108-y}, \href
  {https://doi.org/10.1038/s41467-019-14108-y}
  {\path{doi:10.1038/s41467-019-14108-y}}.

\bibitem{zaidi_international_2019}
Waqar Zaidi and Allan Dafoe.
\newblock International {Control} of {Powerful} {Technology}: {Lessons} from
  the {Baruch} {Plan} for {Nuclear} {Weapons}.
\newblock {\em Working Paper}, 2019.
\newblock URL:
  \url{https://www.fhi.ox.ac.uk/wp-content/uploads/2021/03/International-Control-of-Powerful-Technology-Lessons-from-the-Baruch-Plan-Zaidi-Dafoe-2021.pdf}.

\end{thebibliography}
\end{document}